\def\bq{\begin{equation}}
\def\eq{\end{equation}}
\def\bqa{\begin{eqnarray}}
\def\eqa{\end{eqnarray}}
\def\bqb{\begin{eqnarray*}}
\def\eqb{\end{eqnarray*}}
\documentstyle[12pt]{article}\textwidth 16cm
\def\pr#1#2#3{ Phys. Rev. ${\bf{#1}}$ (#2) #3 }
\def\prl#1#2#3{ Phys. Rev. Lett. ${\bf{#1}}$ (#2) #3 }
\def\pl#1#2#3{ Phys. Lett. ${\bf{#1}}$ (#2) #3 }
\def\prep#1#2#3{ Phys. Reports ${\bf{#1}}$ (#2) #3 }
\def\np#1#2#3{ Nucl. Phys. ${\bf{#1}}$ (#2) #3 }
\def\zp#1#2#3{ Z. Phys. ${\bf{#1}}$ (#2) #3 }

\def\etal{{\it et.al.\/}}

\def\O{ {\cal O }}

\global\nulldelimiterspace = 0pt

\def\roughly#1{\mathrel{\raise.3ex
    \hbox{$#1$\kern-.75em\lower1ex\hbox{$\sim$}}}}
\def\lsim{\roughly<}
\def\gsim{\roughly>}

\def\Uh{\widehat{U}}

\begin{document}
\pagenumbering{arabic}
\thispagestyle{empty}
\def\thefootnote{\fnsymbol{footnote}}
\setcounter{footnote}{1}

\begin{flushright} PM/95-01 \\
January 1995 \\
\end{flushright}
\vspace{2cm}
\begin{center}
{\Large\bf The Bosonic Sector of the Electroweak
Interactions, Status and Tests at Present and Future Colliders}
 \vspace{1.5cm}  \\
{\large  F.M. Renard}
\vspace {0.3cm}  \\
Physique
Math\'{e}matique et Th\'{e}orique,
CNRS-URA 768\\
Universit\'{e} Montpellier II,
 F-34095 Montpellier Cedex 5.\\

\vspace{1.5cm}
{\bf Lectures given at Regensburg University, January 1995}\\
\vspace{0.5cm}
{\bf Extended version of a talk given at the
Festkolloqium Dieter Schildknecht,
Bielefeld, Oct.14th 1994}

\vspace {2cm}

 {\bf Abstract}
\end{center}
\noindent
The status of the Standard Model (SM) is reviewed.
We emphazize the fact
that in spite of the success of the SM for the descrition of
the fermionic sector, the
status of the bosonic sector (gauge and scalar) suffers from many
theoretical deficiencies and from the lack of empirical support.
This situation, which leaves room for several types of extensions or
alternatives to SM, strongly motivates the pursue of
intense efforts for finding hints of New Physics (NP) effects. We
present a phenomenological description valid for energies lying
below the NP scale. We discuss the indirect constraints established
from high precision tests at LEP1, as well as the direct tests that
could be performed at future machines.

\def\thefootnote{\arabic{footnote}}
\setcounter{footnote}{0}
\clearpage
\section{Introduction, the status of the Standard Model}

It is a common leitmotiv to say that
the Standard Model(SM) is largely successful. On the one hand
it is already remarkable that this model is able
to make definite and unambiguous preditions for
all processes involving usual particles.
This property
is the consequence of the gauge principle which allows to predict the
dynamics once a classification group has been chosen. The simpler QED
case with the $U(1)_{EM}$ has been extended to the non-abelian cases
of QCD with $SU(3)_{colour}$ and to the electroweak interactions with
$SU(2)\times U(1)$. However the specific feature of electroweak
interactions is the fact that $W$, $Z$ bosons are massive. The gauge
principle has to be completed with a mass generation mechanism. In the
Standard Model it is chosen as the Higgs mechanism of spontaneous
symmetry breaking (SSB). It is this last property that makes the SM a
renormalizable theory which allows to compute high order effects
and to make the accurate predictions mentioned above.
These predictions practically agree
with all available experimental results. In spite of this success many
questions arise.\par
Let us first quickly review the status of the SM by clearly
separating the
caracteristics of its three sectors:\par
a) The \underline{fermionic sector} contains the constituents of
matter i.e., the three families of
leptons and quarks.\par
b) The \underline{gauge sector} consists in the $\gamma$, $W^{\pm}$, $Z$
and the 8 gluons, as generated by the $SU(3)\times
SU(2)\times U(1)$ gauge group .\par
At this stage both fermions and bosons are massless states. They are
coupled through gauge interactions. Self-boson interactions appear
through the non-abelian Yang-Mills kinetic terms of the $W^{\pm,3}$
gauge bosons.\par
c)The \underline{scalar sector} is constructed with
a complex doublet of Higgs
fields. The gauge couplings of this scalar doublet provide the
$W^{\pm}$ and Z masses proportional to the vacuum expectation value $v$
(the Fermi scale). Fermion masses can be described at the expense of
introducing by hand a set of Yukawa couplings between fermion and Higgs
fields. The Higgs potential generates the Higgs mass and Higgs
self-couplings.\par
The empirical status of these three sectors is now the following. The
fermionic sector is well described by the $SU(2)\times U(1)$
classification with left-handed doublets and right-handed singlets. The
description of their interactions mediated by the gauge bosons, agrees
with the high precision tests performed in particular at LEP1, in some
cases up to a few permille accuracy. This agreement may be
surprizing when one has in mind the broad spectrum of fermions
(from the extremely light neutrinos up to the very heavy top quark),
its peculiarities (special quantum numbers, chiralities, family
replication with a spectacular hierarchy structure,
absence of right-handed neutrinos,
similarities but differences between leptons and quarks).
One could expect to find some
deviations from universality. Maybe the heavy quark sector, not yet
tested at the same accuracy as the light fermion sector, will reveal
some specific features. These questions may seem rather
aesthetical, nevertheless the
proliferation of 90 basic states is a strong motivation for the
search of a simpler and more fundamental theory.
To understand these various points one may need so-called "New
Physics" (NP) structures like the ones
which extend or modify the SM (Technicolour mechanism
(TC), Grand Unified Theories
(GUT), substructures). New concepts like those introduced with
superstrings may also be necessary.
In any case it is not obvious at what
energy scale (between the TeV range and the
Planck mass) these features may originate. \par
The status of the bosonic sector is not yet empirically established
because it is still not possible to perform significant direct tests.
The agreement of the SM predictions with experiments
in fermionic processes (LEP1, low
energy experiments) is often taken as a sign of general validity of the
SM including the bosonic sector, because high order terms
indirectly involve gauge boson
and also higgs boson self-interactions. However as we will see in
Sect.4, these indirect tests first suffer
from a lack of accuracy, but also
from many ambiguities which prevent to give well-defined
model-independent statements. Many extensions of, or alternatives
to, SM are also consistent with the fermionic results.\par
On another hand the bosonic sectors suffer from much more serious
questions and deficiencies. They concern the
origin of three a priori independent gauge
couplings (that one would like
to unify), the origin of SSB, i.e.
the origin of the scalar potential (not generated by the gauge principle
but put by hand), and of the Fermi scale $v$ (the basic mass
scale of the electroweak interactions), the restricted choice of Higgs
doublets, as well as the unpredicted value
of the Higgs mass (seen either as an
unpredicted coupling constant for the $\phi^4$ term or as a new mass
scale).\par
Even more serious are the following two problems concerning the
Higgs sector. One is called "triviality"
and expresses the fact that
the renormalized coupling constant of the  $\phi^4$ term of the Higgs
potential tends to zero when one wants to get rid of the
cut-off introduced for regularization (as one usually does in
renormalizable field theories). This would imply
that SSB disappears in this limit.
The second one is called "the naturalness problem"
and corresponds to the fact
that, at 1-loop, the Higgs mass depends quadratically of the cut-off
and is no longer controlled by the tree level value $M^0_H$. Its value
then only depends on
an outside scale. This is opposite to what happens in
the "natural" fermion mass case where the mass shift is proportional to
the tree level mass term and only weakly (logarithmically) depends on
the cut-off. This seems to indicate that the description of
the scalar sector of the SM is
not in a fundamental stage but must be considered as an
\underline{effective} one, valid below a certain NP scale $\Lambda$
(identified with the cut-off).\par
 It is the two facts, absence of direct
tests and existence of hard questions, which render further
studies of the
bosonic sectors of primary importance. In the next Sect.2 we list
in more details the precise pragmatical questions to be asked about
$W^{\pm}$, $Z$, $\gamma$ and $H$ properties. In Sect.3 we present
phenomenological tools which should allow to describe these
properties in a
rather model-independent way. Sect.4 is devoted to the discussion of
the indirect tests performed at LEP1 and Sect.5 to the direct tests
that can be achieved at future machines.
Some perspectives
are outlined in Sect.6.\par

\newpage
\section{Questions to be asked about the bosonic sector}
\par

The $W^{\pm}$ and $Z$ bosons have been discovered in a range
of mass which
precisely agrees with the one
expected from the properties of the weak interactions found in
low energy experiments. The high precision tests which followed
their discovery have confirmed that their couplings to leptons and
quarks agree up to a few permille with the SM predictions \cite{LEP1}.
Can one from
that conclude that $W$ and $Z$ have exactly the gauge nature that the
SM assumes for them? Does it also mean that the Higgs mechanism is
necessarily responsible for mass generation?\par
Certainly not! In fact many options for non-standard (NP) models are
still allowed by the presently limited
empirical knowledge and one can ask the
following questions, classified into three types.\par
\vspace{0.1cm}
{\bf a) The nature of the $W^{\pm}$, $Z$ bosons.}\par
Are they true gauge bosons? In that case what is the precise gauge
group? $SU(2)\times U(1)$ or a larger one like $SU(2)\times SU(2)\times
U(1)$ or $SU(2)\times  U(1) \times  U(1)$? Such extensions
like Left-Right symmetry, E6 symmetry \cite{revE6} are obtained
on the way
of a Grand Unified Theory (GUT) \cite{revGUT}
or in certain
alternative mass generation mechanisms based on a strongly interacting
sector \cite{BESS},\cite{KKS}.\par
More drastically departing from the SM picture,
$W^{\pm}$ and $Z$ may be kind of massive vector states
(hadron like) whose interactions respect some global symmetry. This is
what happens in compositeness schemes where the global symmetry
originates from the subconstituent structure.
This ensures that the
couplings to leptons and quarks are similar to the SM ones\cite{comp}.
Mass may here simply
originate from confinement effects.\par
\vspace{0.1cm}
{\bf b) The precise spectrum of weak bosons.}\par
\underline{Vector bosons}\par
Are there higher vector bosons? They could be either additional gauge
bosons associated to an extended gauge group ($W^{\pm}_R$, $Z_R$, $Z'$,
$V^{\pm,0}$,...)\cite{revE6},\cite{BESS},\cite{KKS}
or partners of  $W^{\pm}$ and $Z$,
like isoscalar vector bosons ($Y$, $Y_L$,...) or excited states
($W^{\pm*}$, $Z^*$), in alternative (for
ex. composite)
schemes\cite{comp}.\par
\underline{Scalar bosons}.\par
Does the Higgs boson exist at all? This question arises because
there exist alternative models
without Higgs (Technicolour-like \cite{TC}or compositeness
inspired\cite{comp}).
If the Higgs exists, is it an elementary or a
composite state \cite{TC}? .
If it exists as
an elementary state, mainly because of the
naturalness problem\cite{Haber}, the question arises whether
it is light (close to $M_Z$) or heavy
(close to the unitarity limit in the TeV range). If it is light, is it
accompanied by other neutral $H^0$ states and charged $H^{\pm}$ states
(as claimed by Supersymmetry in order to cancel the quadratic
divergences) \cite{Haber}? \par
One can also raise the question weather
there exist higher spin ($J \geq 2$) bosonic states?\par
\vspace{0.1cm}
{\bf c) The precise structure of the bosonic interactions?}\par
This question is motivated by the fact that any extension or
modification of the SM should lead to "anomalous" interactions among
usual bosons. In the vector boson subsector, the basic $W$, $Z$,
$\gamma$ self-interactions can be different from the Yang-Mills ones.
In particular new forms and new multi-boson
interactions could appear.
Couplings involving longitudinal $W_L$ states may have
special features related to the fact that they are created by the
mass generation mechanism (MGM). This feature is a genuine one as
compared to the QED or QCD cases where SSB does not occur.
Within the SM structure  it is already
known that a very heavy Higgs is a
source of strong $W_L W_L$ interactions \cite{chan}. New Physics
structures may also introduce further differences between $W_T$ and $W_L$
interactions \cite{GR2}.\par
Obviously the Higgs sector should be directly affected by the existence
of a different MGM, especially Higgs self-interactions because
they reflect the
structure of the potential. Scalar boson-Vector boson couplings
would also be modified if the origin of the scalar boson is non
standard, for example like in TC\cite{TC} or in any other
compositeness schemes\cite{comp}.\par
In order to answer these questions, precision tests of the bosonic
sectors (gauge and scalar) have to be performed. Because of the rich
variety of possible NP schemes, the analyses of
present and future experiments must be done in the most possible
unbiased and model independent way. This is the aim of the
phenomenological description presented in the next Section.\par

\newpage
\section{Phenomenological description of the bosonic sector}
\par

Searches for NP effects can be divided into 2 classes:\par
(A) search for
new particles which cannot fit into the SM classification (not a new
family of leptons and quarks, not a Higgs scalar), and \par
(B) search for
anomalous interactions among usual particles due to residual effects of
NP.\par
 In both cases we can look for direct as well as for indirect
effects of these new particles or interactions. The characteristic
scale of NP is generally expected to lie in the TeV range (following
arguments based on unitarity, on the TC mechanism or simply on present
experimental limits). If this is true, then new particles should more
probably have masses in this TeV range so that their direct
production requires high energy colliders. It is however not
excluded that some states have lower masses and can be found
earlier. If this is not the case one can nevertheless indirectly
try, from their virtual effects
in certain processes (mixing effects with
usual particles, effects through loop diagrams),
to find hints of their existence.
Similarly the existence of new interactions can be directly observed in
processes involving gauge bosons and Higgs bosons. But they could also
be detected through indirect effects
in fermionic processes (like loops involving self-boson
couplings), measured with a very high accuracy as
it is the case at Z peak.\par
\vspace{0.1cm}
{\bf A1) Direct production of new particles}\par
The rate for new particle production in a collider is
essentially controlled by the product  $\sigma \times B$ of the
production cross section times the branching ratio of the new particle
decay mode into the channel that is detected. When no candidate event is
observed a mass limit for the new particle is given. This is
significant only if the coupling of the new particle to the initial
and to the final states consisting of usual particles
is sufficiently strong so that
$\sigma \times B$
reaches the observability limit of the experiment. This is a very model
dependent question and it explains why mass limits given in the
literature are so strongly process dependent and why
the results are so largely spred out.
As one essentially uses fermionic processes, limits appear
to be especially low for those states that are weakly coupled to
usual leptons or quarks, i.e. $M_H \geq 60 GeV$ from LEP1
\cite{Higgs}, $M_V \geq
250 GeV$ for the V bosons generated by the strongly
interacting sector \cite{BESS},\cite{KKS}.
On the opposite,  in other cases they
approach the TeV range\cite{FNAL}.
The low values quoted above illustrate the
fact that indeed, at present, the bosonic sector is
still very weakly constrained.\par
\vspace{0.1cm}

{\bf A2) Indirect effects of new particles}\par

As an example of indirect effect of heavier particles we shall treat
the $Z-Z'$ mixing case which has been extensively studied at LEP1
\cite{Zprime}. We
shall first present a rather general model-independent description and
then look at specific models.\par
If the $Z^0$ mixes with a higher $Z'^0$  vector boson with a mixing
angle $\theta_M$
\bq   Z = Z^0cos\theta_M + Z'^0sin\theta_M  \eq
its vector $g_{Vf}$ and axial $g_{Af}$ couplings get modified as follows
\bq   \delta g_{Vf} = G'\theta_M{c_f+d_f\over2} \ \ \ \ \
\delta g_{Af} = G'\theta_M{c_f-d_f\over2}  \eq
depending on the $Z'^0f \bar f$ couplings defined as
\bq   -i{eG'\over4sc}\gamma^{\mu}[{1-\gamma^5\over2}c_f +
{1+\gamma^5\over2}d_f]\eq
{}From eq.(2) one sees that the description will involve 7 independent
parameters ($c_f$ and $d_f$) when one assumes family universality,
i.e. $f_{L,R}$ respresenting $\nu_L,
l_{L,R}, u_{L,R}, d_{L,R}$ states.
In $Z$ peak experiments the disentangling
of these 7 parameters will require the largest set of observables.\par
For the three parameters of the leptonic sector
one has the three following
observables:
the charged leptonic Z partial width  $\Gamma_l$, the neutral one
$\Gamma_{\nu}$, and the leptonic
asymmetry $A_l$, defined as $A_{LR}$, but also
measurable through the tau lepton
final polarization asymmetry or through the
forward-backward asymmetry $A_{FB,l} = {3\over4}A^2_l$.\par
For the 4 parameters of the quark sector one can
take the following two partial
widths
 $\Gamma_4=\Gamma_u+\Gamma_d+\Gamma_c+\Gamma_s$ ,\ \
$\Gamma_b$
and the two asymmetries
  $ A_c$,\ \ $ A_b$.  \par
Only $\Gamma_4$ is presently available with a high accuracy. It is
in fact more convenient \cite{Zprime} to use the combination \cite{D}
\bq  D = {\Gamma_4\over\Gamma_l} - 2(3-{20\over3}s^2)A_l   \eq
The forward-backward asymmetries
$A_{FB q}={3\over4}A_lA_q$ are not accurate
enough to determine $A_q$,
so that one needs measurements of the
polarized asymmetries $A^{pol,q}_{FB}={3\over4}A_q$
for $q=c,b$ in order to get
a meaningful
result \cite{Zprime},\cite{pol}, \cite{blrv}.\par

\underline{Application to specific models}\par
Various types of extensions of the SM (like $E_6$ or $L-R$ symmetry) or
of alternative models can be treated in this manner. In each specific
case, $c_f$ and $d_f$ are fixed by the classification group and in
some cases $G'$ is
related to the electroweak strength by unification conditions. The only
free parameter is then $\theta_M$ which can also be related to the mass
ratio ${M_Z \over M_Z'}$. In Sect.4  we will see how LEP1 results allow
to give upper limits for $\theta_M$ and hence to give lower mass limits
for the $Z'$.\par
\vspace{0.1cm}

{\bf B) Residual bosonic interactions below New Physics threshold.}\par

We now present the description of residual interactions among usual
particles. We anticipate the discussion of results
from Z peak physics which strongly
constrain (at the permille level)all non SM effects involving light
fermions. We restrict to couplings involving $W^{\pm}$,
$Z$, $\gamma$ and
Higgses, avoiding those which involve lepton and quark fields.
The case of couplings involving a heavy top quark is still an
opened question which is under study\cite{Otop}.
Let us start by recalling the
basic SM bosonic couplings.\par
\underline{SM self-couplings at tree level}\par
In the SM, self-gauge boson couplings are
given by the Yang-Mills structure of the
kinetic terms of
the W triplet (no B field is involved). It produces 3-boson and 4-boson
couplings involving at least one $W^+W^-$ pair (no pure neutral
coupling exist).
\bq L_W=-{1\over2}<W_{\mu \nu}W^{\mu \nu}>
\eq

Higgs boson couplings with gauge bosons are given by the covariant
derivative of the scalar kinetic terms.
\bq
L_{\Phi}=(D_{\mu}\Phi^+)(D^{\mu}\Phi)={v^2\over2}<D_{\mu}UD^{\mu}U^+> \eq
Three- and four-Higgs couplings are given by the potential term
\bq L_V=-V=-{M^2_H\over2v^2}(\Phi^+\Phi-{v^2\over2})^2=C+\mu^2\Phi^+\Phi
+\lambda(\Phi^+\Phi)^2\eq
\bq C={M^2_Hv^2\over2} \ \ \ \ \mu^2=-{M^2_H\over2}\ \ \ \
v^2=-{\mu^2\over\lambda} \eq
Our notations are the following ones:
\bq {W}^a_{\mu \nu}=\partial_{\mu}W^a_{\nu}-\partial_{\nu}W^a_{\mu}
-g\epsilon^{abc}W^b_{\mu}W^c_{\mu} \eq
\bq
W_\mu = \overrightarrow{W}_\mu \cdot
\frac{\overrightarrow{\tau}}{2} \ \ \ \ \ , \ \ \ \ \ \ \
W_{\mu \nu}= \overrightarrow{W}_{\mu \nu} \cdot
\frac{\overrightarrow{\tau}}{2} \ \ \ \ \ , \ \
\eq

\bq \Phi=\left( \begin{array}{c}
      \phi^+ \\
{1\over\sqrt2}(v+H+i\phi^0) \end{array} \right) \ \ \ \ , \ \
\eq
\bqa
D_{\mu} & = & (\partial_\mu + i~ g_1 Y B_\mu +
i~ g_2 W_\mu ) \ \ \ \ , \ \ \nonumber
\eqa

\bq
\Uh={v\over\sqrt2}U=\bigm(\widetilde \Phi\ \ , \ \Phi\bigm) \ \ \ \
, \eq
where $\widetilde \Phi = i\tau_2 \Phi^* $
and $\langle A \rangle \equiv TrA$.

\par

\underline{Standard radiative corrections}, at 1-loop (fermion and
boson loops) generate form factors associated to each of the SM
tree level terms but also new coupling forms which do not exist at tree
level\cite{corWW}. We shall illustrate the case of 3-gauge boson couplings
(analogous studies have been done for 4-boson and for Higgs
couplings).\par
\newpage
\underline{General Lorentz and U(1) invariant
forms for $ZW^+W^-$ and $\gamma
W^+W^-$ couplings}\par
The complete set has been established in \cite{GG}. It involves seven
independent $VW^+W^-$ forms for both $V=Z, \gamma$
which are listed below:\par
\noindent
1)
\bq \ \ \ \ \ -ieg_VV_{\mu}[\widetilde W^{-\mu\nu}W^+_{\nu}
-\widetilde W^{+\mu\nu}W^-_{\nu}] \eq
2)
\bq \ \ \ \ \  -ieg_V\kappa_V\widetilde V_{\mu\nu}W^{+\mu}W^{-\nu}
\eq
3)
\bq \ \ \ \ \  +ieg^{SM}_V{\lambda_V\over
M^2_W}\widetilde V_{\nu\lambda}\widetilde W^{-\lambda\mu}
\widetilde W^{+\mu}_{\nu} \eq
4)
\bq \ \ \ \ \ {ez_V\over
M^2_W}\partial_{\alpha}\hat Z_{\rho\sigma}
(\partial^{\rho}W^{-\sigma}W^{+\alpha}
-\partial^{\rho}W^{-\alpha}W^{+\sigma}+\partial^{\rho}W^{+\sigma}W^{-
\alpha}-\partial^{\rho}W^{+\alpha}W^{-\sigma}) \eq
5)
\bq \ \ \ \ \ ieg^{SM}_V\hat\kappa_V\hat
 Z_{\mu\nu}W^{+\mu}W^{-\nu} \eq
6)
\bq \ \ \ \ \ {\hat\lambda_V\over
M^2_W}\hat Z^{\nu\lambda}\widetilde W^+_{\lambda\mu}
\widetilde W^{-\mu}_{\nu} \eq
7)
 \bq \ \ \ \ \ eg^{SM}_VK_V(\partial^{\mu}Z^{\nu}+
\partial^{\nu}Z^{\mu})
W^+_{\mu}W^-_{\nu} \eq

where the abelian $\widetilde W^a_{\mu \nu}
=\partial_{\mu}W^a_{\nu}-\partial_{\nu}W^a_{\mu}$ is used as well as
the dual
\bq \hat Z_{\mu\nu}={1\over2}
\epsilon_{\mu\nu\alpha\beta}(\partial_{\alpha}
Z_{\beta}-\partial_{\beta}Z_{\alpha}) \eq
In ref. \cite{BMTnlc} and \cite{BMTlep2} the following combinations of
couplings are defined
\bq \delta_V=g_V-g^{SM}_V \ \ \ \
 x_V=(\kappa_V-1)g_V \ \ \ \
 y_V=\lambda_Vg^{SM}_V \eq
\bq z'_{1V}=g^{SM}_VK_V \ \ \ \
 z'_{2V}=g^{SM}_V(\hat\kappa_V-\hat\lambda_V) \ \ \ \
  z'_{3V}=g^{SM}_V\hat\lambda_V/2  \eq

The SM case corresponds to\par

\bq g^{SM}_{\gamma}=1 \eq
\bq  g^{SM}_Z=cot\theta_W \eq

and all other terms being absent.\par
\newpage

As summarized in Table 1 the three first terms are C- and P- conserving
(charge,
magnetic moment and quadrupole moment), the fourth one is
C- and P- violating but CP-
conserving (anapole term) and the last three ones are
CP- violating. The specific
helicity properties \cite{BMTnlc},\cite{BMTlep2} of the $W^+W^-$ state
for each type of coupling are also given in the last three
lines of Table 1. The identification of
these properties is particularly useful
for experimental analysis as it gives a way
to disentangle the
various forms.

\begin{center}
\underline{Table 1: Space-time properties of the seven
3-boson coupling forms}\par
\vspace{0.2cm}

\begin{tabular}{|c|c|c|c|c|c|c|} \hline
\multicolumn{1}{|c|}{$\delta_V$} &
  \multicolumn{1}{|c|}{$x_V$} &
   \multicolumn{1}{|c|}{$y_V$} &
    \multicolumn{1}{|c|}{$z_V$} &
     \multicolumn{1}{|c|}{$z'_{1V}$} &
      \multicolumn{1}{|c|}{$z'_{2V}$} &
       \multicolumn{1}{|c|}{$z'_{3V}$}  \\[.1cm] \hline
  P & P & P & & P & &  \\ \hline
 C & C & C & &  & C & C  \\ \hline
 CP & CP & CP & CP &  &  &   \\ \hline
 TT &  & TT &  &  & TT & TT  \\ \hline
 LL & LL &  &  &  &  &   \\ \hline
 LT & LT & LT  & LT & LT & LT &   \\ \hline
\end{tabular}
\end{center}
\noindent

 SM radiative corrections feed all these terms with
$q^2$-dependent form factors.\par
\underline{Departures due to NP}\par
NP can contribute to such new
couplings and form factors. This may happen in various ways. The basic
$W$,$Z$ structure may differ from the SM one if one uses an alternative
description, for ex. if $W$,$Z$ are massive vector bosons not directly
generated by the gauge principle (composite states like hadronic
$\rho$, $\omega$,...vector mesons)\cite{KMSS}, \cite{Grosse}.
In these cases tree level
modifications of the self-boson couplings (finite $\delta \kappa$,
$\lambda$,...) may exist. In less drastic pictures in which the
$SU(2)\times U(1)$ system is kept but extended or coupled to a new
additional sector, tree level modifications may still
appear through mixing
of $W$, $Z$ with higher vector bosons (especially if these ones pertain
to a strongly interacting sector like $SU(2)_V$)\cite{BESS},
\cite{KKS}.
In any case
at 1-loop, NP effects
will always appear through contributions of virtual states. They can
even be enhanced by non-perturbative effects(hypercolour factors,
resonant effects,...). The peculiarities of the
terms generated in this way \cite{GRdyn}
(for example the specific sectors that they affect,
charged versus neutral states, transverse versus longitudinal ones,
Higgs versus no-Higgs final states,...) and
the symmetries that they respect should reflect their origin and help to
identify the nature of NP through
detailed analyses of the processes.\par
We shall discuss these questions in a precise manner through
the effective
lagrangian method. If the characteristic scale $\Lambda$ of NP
is sufficiently
larger than $M_W$, effective lagrangians among usual particles are
obtained by integrating out all heavy degrees of freedom. They can
be written in the form
\bq   L = \Sigma_i {\bar f_i \over \Lambda^{d-4}}O^{(d)}_i  \eq
in which $O^{(d)}_i$ are operators of dimension $d$
constructed with usual fields,
$\Lambda^{d-4}$ is a scale factor ensuring that $L$ has the correct
dimension 4 when the coupling constants $f_i$ are dimensionless.
A priori
such a series can be infinite and one needs restrictions in order to
have in practice a useful description. These restrictions must be done
on a physical basis because often an apparently "harmless" mathematical
property can have very important physical consequences. As already said
and motivated by LEP1 results we restrict $O^i$ to not involve lepton
and quark fields. The next restriction comes from the
dimension. If
$\Lambda >> M_W$ it is natural to expect observable effects only from
the lowest dimensions $d= 4,6$, perhaps 8.\par
\underline{Global symmetries}\par
The above method leads to a large set of possible coupling
forms. One can try to reduce this list by
demanding that the lagrangian satisfy certain global symmetries which
are empirically known to be essential. For example the
global $SU(2)_{Weak}$
symmetry ensures the correct form of the W-fermion
couplings\cite{KMSS}. Broken by electromagnetism
through $\gamma-W^3$ junction,
after mixing,
it produces the physical photon and the physical $Z$. In
this picture if one considers the Lorentz invariant, $U(1)_{EM}$
invariant and global $SU(2)_{Weak}$ invariant, d=4 forms constructed
without Higgs,
one obtains a set of couplings involving four free parameters
(two for the
3-boson part and two for the 4-boson one)\cite{KMSS}.
The two parameters of the
3-boson part contribute to the departure of the
Yang-Mills $ZW^+W^-$ coupling constant from the SM value, $\delta_Z =
g_Z-cot\theta_W$
and to the anomalous magnetic moment
couplings $ZW^+W^-$ and $\gamma W^+W^-$ satisfying the relation
\bq  x_Z = -{s\over c}x_{\gamma}  \eq
No quadrupole coupling is generated at this level. This set of free
parameters can be further reduced if one considers the
high energy behaviour of boson-boson
scattering amplitudes. Because of these non-standard terms,
they grow like $s^2$.
Demanding that these terms cancel, one obtains
 certain relations among the
four free parameters which finally reduce to only one\cite{BKS}.
\bq x_Z = -{s\over c}x_{\gamma} = -s^2\delta_Z \eq
The amplitudes then only grow like s. This is the level at which Higgs
contributions would appear. We shall come back to this point later
on.\par
If one wants to generate a quadrupole coupling, d=6 terms have to be
allowed. Only
one free parameter is generated if one demands the
exact validity of the global $SU(2)_L$ symmetry\cite{KRS}, so that one
obtains
\bq   \lambda_{\gamma} = \lambda_Z \eq
This requirement of global $SU(2)_L$ is motivated by the concept of
custodial symmetry\cite{Sikivie}. It is this concept which
explains why, in spite of
the large symmetry breaking (for ex. $m_t >> m_b$) the ratio $\rho$
remains very close to one (up to radiative correction effect whose
leading term is ${\alpha \over \pi}{m^2_t\over M^2_Z}$). This means that
there is essentially no violation of the $W$ triplet structure (i.e.
$SU(2)_L$ global symmetry). The custodial symmetry can be
defined as any
$SU(2)_c$ global symmetry under which the W fields
transforms as a triplet.
Before SSB the scalar sector of the SM satisfies a
global $SU(2)_L \times SU(2)_R$ symmetry. SSB breaks
this symmetry down to
$SU(2)_c$ and this is what ensures $\rho = 1$. Note that $SU(2)_c$
is broken by the gauge coupling of the B field. This is why it is often
discussed only in the limit $g'\to 0$.\par

\underline{Local symmetries}\par
The simplest case that we shall develop
here is the one in which the $SU(2)
\times U(1)$ SM structure is extended by group factors whose degrees of
freedom are associated to a heavy NP scale and will be integrated out.
When SSB occurs the effective lagrangian
$L(W, B, \Phi)$ which satisfies local $SU(2)\times
U(1)$ produces the effective $L(W, Z, \gamma, H)$.\par
 There are other
possibilities. For example the basic group $SU(2)\times SU(2)\times
U(1)$ \cite{KKS}
can be directly broken to $U(1)_{EM}$ without passing through
$SU(2)\times U(1)$. In such cases the effective lagrangian explicitely
breaks $SU(2) \times U(1)$ gauge invariance. Another example is the one
in which the Higgs mass is very large so that
the Higgs field practically
disappears from the spectrum, leading also to breaking of $SU(2) \times
U(1)$ gauge invariance.\par For simplicity and also because
$SU(2)\times U(1)$ gauge invariance ensures many good properties for
high order effects\cite{DeR}, we shall concentrate on this gauge
invariant case.
However one must
realize that this requirement of gauge invariance by itself does not
restrict the number of independent couplings. It is always possible to
find a suitable combination of scalar field which render gauge
invariant a given
self-boson coupling written in the unitary gauge\cite{numb}. It
is only when one simultaneously restrict the dimension that one gets an
appreciable reduction of the number of free parameters. For $d=4$ all
the $SU(2)\times U(1)$ gauge invariant terms are already provided by
SM. Only at $d=6$ start the new effective terms. In terms of pure
bosonic fields there are 11 independent
CP-conserving such terms\cite{Buchmuller}, \cite{Hag}.
They are listed below.\par

\bqa
\overline{\O}_{DW} & =&  4 ~ \langle ([D_{\mu} ,
W^{\mu \rho}])([D^{\nu} ,
W_{\nu \rho}]) \rangle \ \ \
\ \ \ \ \ \ \ \ \ \  , \ \ \ \\[0.3cm]
\O_{DB} & = & (\partial_{\mu}B_{\nu \rho})(\partial^\mu B^{\nu
\rho}) \ \ \ \ \ \ \ \ \ \ \ \ \ \ \ \ \ \ \ \ \ \ \
\ , \ \ \ \\[0.3cm]
\O_{BW} & =& \Phi^\dagger B_{\mu \nu} W^{\mu \nu} \Phi
\ \ \ \ \ \ \ \ \ \ \ \ \ \ \ \ \ \ \ \ \ \
 \ \ \ \ \ \ \ , \ \ \\[0.3cm]
\O_{\Phi 1} & =& (D_\mu \Phi)^\dagger \Phi \Phi^\dagger
(D^\mu \Phi) \ \ \ \ \ \ \ \ \ \ \ \ \ \ \ \ \ \ \ \ \ . \ \ \
\eqa

\bqa
\O_W &= & {1\over3!}\left( \overrightarrow{W}^{\ \ \nu}_\mu\times
  \overrightarrow{W}^{\ \ \lambda}_\nu \right) \cdot
  \overrightarrow{W}^{\ \ \mu}_\lambda =-{2i\over3}
\langle W^{\nu\lambda}W_{\lambda\mu}W^\mu_{\ \ \nu}\rangle \ \ \
, \ \ \ \ \\[0.3cm]
\widehat{\O}_{UW} & = & \frac{1}{2}\, \langle \Uh\Uh^{\dagger}
\rangle  \langle W^{\mu\nu} \
W_{\mu\nu}\rangle \ \ \ \ \ \ \ \ , \ \ \ \\[0.3cm]
\widehat{\O}_{UB} & = & \langle \Uh\Uh^{\dagger} \rangle B^{\mu\nu} \
B_{\mu\nu} \ \ \ \ \ \ \ \ \ , \ \ \ \\[0.3cm]
\O_{W\Phi} & = & 2~ (D_\mu \Phi)^\dagger W^{\mu \nu} (D_\nu \Phi) \ \ \ \ \
\ \ \ \ \ , \ \  \ \\[0.3cm]
\O_{B\Phi} & = & (D_\mu \Phi)^\dagger B^{\mu \nu} (D_\nu \Phi)\ \ \ \ \
\ \ \ \ \ \ , \ \  \
\eqa

\bqa
\O_{\Phi 2} & = & ( \partial_\mu \langle \Uh \Uh^\dagger \rangle )
(\partial^\mu \langle \Uh \Uh^\dagger \rangle ) \ \ \ \ \ \ \ , \ \ \
\\[0.3cm]
\O_{\Phi 3} & = &  \langle \Uh \Uh^\dagger \rangle ^3\ \ \ \ \ \ . \ \
\eqa

As presented in Table 2, one can regroup the operators into sets
which have basically
different physical consequences and behaviours under certain
symmetries. The first four of them are called non-blind\cite{DeR}
because they involve
2-point gauge boson functions. They would then
directly affect the observables
measured at LEP1. Consequently their coupling constants must have
a strongly reduced strength in order to avoid direct observation.
The next five ones are the "blind" ones in
the sense that LEP1 is blind to them at tree level. They can only
affect the LEP1 observables through 1-loop. The resulting
constraints are
very mild and allow for large values of the coupling constants. The
last two ones only involve Higgs fields and has been dubbed
"super-blind"\cite{GRdyn}
because they are almost unconstrainable by present and
future machines.\par
%
\begin{center}
\underline{Table 2:  Properties of the eleven bosonic operators}\par
\vspace{0.2cm}

\begin{tabular}{|c|c|c|c|c|} \hline
\multicolumn{1}{|c|}{operator} &
 \multicolumn{1}{|c|}{non-blind} &
  \multicolumn{1}{|c|}{blind} &
   \multicolumn{1}{|c|}{super-blind} &
    \multicolumn{1}{|c|}{$SU(2)_c$} \\[.1cm] \hline
$\overline{\O}_{DW}$  & x &  &  & x    \\ \hline
$\O_{DB}$ & x &  &  &     \\ \hline
$\O_{BW}$ & x &  &  &     \\ \hline
$\O_{\Phi 1}$ & x &  &  &     \\ \hline
$\O_W$ &  & x &  & x    \\ \hline
$\widehat{\O}_{UW}$ &  & x &  & x    \\ \hline
$\widehat{\O}_{UB}$ &  & x &  &     \\ \hline
$\O_{W\Phi}$ &  & x &  &     \\ \hline
$\O_{B\Phi}$ &  & x &  &     \\ \hline
$\O_{\Phi 2}$ &  &  & x & x    \\ \hline
$\O_{\Phi 3}$ &  &  & x & x   \\ \hline
\end{tabular}
\end{center}
\noindent

Let us concentrate on the 5 blind ones. It is interesting to examine
how they contribute to self-boson couplings.\par
$\O_W$ leads to the famous
quadrupole type\cite{KRS} of coupling
with the relation given in eq.(28) .\par
$\O_{W\Phi}$ and $\O_{B\Phi}$ contribute to
$\delta_Z$, $\kappa_{\gamma}$
and $\kappa_Z$, satisfying the relations(26) previously obtained in the
general $SU(2)_W$ invariant schemes\cite{KMSS}, \cite{Grosse}.
In particular  $\O_{W\Phi}$
reproduces the special case \cite{BKS} satisfying relations (27).\par
The two other
operators $\widehat \O_{UW}$ and $\widehat \O_{UB}$
only contribute to anomalous
Higgs-gauge boson couplings\cite{GRdyn}.\par

\underline{Custodial symmetry}\par
Demanding a strict application of the custodial symmetry for the NP
effects, strongly restricts the list of operators,
see Table 2. From the 11 above
ones only five of them are $SU(2)_c$ invariant namely, $\bar O_{DW}$,
$O_W$, $\widehat \O_{UW}$ and the two
superblind $\O_{\phi2}$ and $\O_{\Phi3}$.
$SU(2)_c$ symmetry restricts the 5 blind ones to only two.
Remember that one of them, $\O_W$ was already obtained from the
$SU(2)_L$ global symmetry which is  for the pure
W sector, a remnant  of the full
$SU(2)_c$. The other one is $\O_{UW}$
which also involves Higgs fields. The justification for this strict
use of custodial symmetry is that NP is supposed to be intimately
related to the origin of the scalar sector and should therefore
respects the same symmetries.\par

 \underline{Chiral descriptions}\par

Let us consider a situation in which the $SU(2)\times U(1)$ electroweak
symmetry is broken by a strongly interacting sector, all new particles
including the Higgs boson being much heavier than $M_W$. In such a
situation it is convenient to use a non-linear representation in which
the $U$ matrix of eq.(12) containing the three goldstone degrees of
freedom is written as
\bq  U=e^{i{\overrightarrow{\xi}.\overrightarrow\tau\over v}}  \eq

Effective lagrangians invariant under $SU(2)\times U(1)$
resulting from integrating out the effects of
this sector can be constructed
as combinations of gauge boson fields, $U$ matrices and
their covariant derivatives. At present energies it is meaningful
to make an expansion with respect to the number p of
derivatives or of gauge fields ($U$ being dimensionless). At lowest
($p^2$) order one finds the SM part eq.(6). New
couplings appear at order $p^4$, $p^6$,...etc. In this way one can again
generate all possible bosonic operators. In the physical gauge, they
produce the set of anomalous 3-boson couplings listed
above as well as higher multi-boson couplings. However the difference
with the linear representation presented before is the absence of a
physical Higgs field and a different ordering in magnitude
of the anomalous
self-boson couplings. For
example $\delta_Z$, $\kappa_{\gamma}$ and  $\kappa_Z$ appear at order
$p^4$, through the operators called $L_{9L}$ and $L_{9R}$ and satisfy
eq.(26)
\bq   L= -igL_{9L} \langle W^{\mu\nu} D_{\mu}U
D_{\nu}U^{\dagger}\rangle   -ig'L_{9R} \langle B^{\mu\nu} D_{\mu}
U D_{\nu}U^{\dagger}\rangle \eq
\bq  \delta_Z={e^2\over 2cs^3}L_{9L} \eq
\bq   x_Z=-{s\over c}x_{\gamma}=-{e^2\over2cs}(L_{9L}+L_{9R})   \eq

On another hand the quadrupole coupling
$\lambda$ appears at order $p^6$ through $L_{\lambda}$ which
is just the usual operator $O_W$ of eq.(33). For more details and
specific applications see ref \cite{chiral}.

\underline{Unitarity constraints}.\par
When operators with $d > 4$ are considered, they generally lead to
boson-boson scattering amplitudes which grow fastly with the center of
mass energy. For example $d=6$ terms lead to partial wave amplitudes
growing like $s$ or $s^2$. This means that for a given value of the
coupling constant the amplitudes reach the unitarity limit at a certain
energy scale. At this point unitarity saturation effects (resonances or
new particle creation,...) must occur. So the unitarity relations which
are obtained for each of the operators have two meanings.\par
1. For a given coupling constant one obtains a value for the scale at
which unitarity saturation occurs (this can be considered as a practical
definition of the NP scale),\par
2. For a given NP scale one can set upper limits for the coupling
constants in order to satisfy unitarity in the whole $s \leq
\Lambda^2$ domain.\par
For the 5 blind operators the unitarity constraints read\cite{GLRuni},
\cite{GLRUNI}

\bq
|f_B| \leq 98{M^2_W\over{s}} \ \ \ \ \ , \ \ \ \ \
 \ |f_W| \leq 31{M^2_W \over{s}} \ \ \ \ ,\ \ \ \ \
|\lambda_W| \lsim 19~{M^2_W \over s} \, \eq

\bq
|\lambda_W| \lsim 19~{M^2_W \over s}  \eq

\bq  |d| \lsim 17.6~{M^2_W\over{s}}+2.43
{M_W\over{\sqrt{s}}}
\eq

\bq
-236~{M^2_W\over{s}}~+~1070~{M^3_W\over{s^{3/2}}}~ \lsim ~d_B ~\lsim
{}~ 192~{M^2_W\over{s}}~-~
1123~{M^3_W\over{s^{3/2}}}  \ \ \ \ .\ \ \
\eq

If one fixes the NP scale at 1 TeV, the coupling constants have to
satisfy the bounds
\bq
|f_B| \lsim 0.6 \ \ \ , \ \ \ |f_W| \leq 0.2 \ \ \,\ \ \ \
  |\lambda_W| \lsim 0.12   \eq

\bq |d| \lsim 0.3  \ \ \, \ \ \  -0.8 \lsim d_B \lsim 0.6
\ \ \ \ . \ \ \eq

We shall see in the next Sect.4 that these bounds are highly
non trivial as
compared to the indirect constraints obtained from LEP1. On the
opposite, similar bounds obtained for non-blind operators are totally
useless as they lie far above the very stringent LEP1 limits.\par

\newpage
\section{Status after the high precision tests at LEP 1}\par

It is interesting to discuss how far the high precision tests done at Z
peak with fermionic processes can be used to test the bosonic sector.
In order to achieve this goal it is essential to use a description of
the Z exchange processes which is sufficiently general in order to
account for possible NP effects but also to cover in an accurate way the
SM radiative correction effects ($W$, $Z$ self-energies, vertex and box
corrections). For this reason  the usual description \cite{AB},
\cite{Hollik} of the effective
$Z$ exchange amplitude in $e^+e^- \to f \bar f$ has been
somewhat generalized \cite{vertex}.\par
{\bf A) Formalism}\par
 We write it in the form
$$ A^Z={\sqrt{2}G_{\mu}M^2_Z\over
q^2-M^2_Z-iM_Z\Gamma_Z(q^2)}[1+\delta^{s.e.}][\bar
v_e\gamma^{\mu}([g_{Ve}+\Delta g_{Ve}]-\gamma^5[g_{Ae}+\Delta
g_{Ae}])u_e]\times  $$
\bq\times[\bar
u_f\gamma^{\mu}([g_{Vf}+\Delta g_{Vf}]-\gamma^5[g_{Af}+\Delta
g_{Af}])v_f]   \eq

The SM part at 1-loop is fully taken into account through the
three inputs $\alpha(0)$, $G_{\mu}$, $M_Z$, and through the shifts
$\delta^{s.e.}$, $\Delta g_{Vf}$ and $\Delta g_{Af}$.
{}From the inputs one derives
\bq s^2_1c^2_1 = {\pi \alpha(0)\over\sqrt{2} G_{\mu}M^2_Z}  \eq
and the basic $g_{Vf}$ and $g_{Af}$ couplings
\bq g_{Vf}=I^3_f-2s^2_1Q_f   \ , \ \ g_{Af}=I^3_f \ .\eq
The shifts
 contain the SM radiative correction effects (in particular the
 large $m_t$ and $M_H$ dependent terms) and the NP contributions.
 We have already seen in Sect.3
how $Z-Z'$ mixing effects modify the
Z couplings, i.e. add $\delta g_{Vf}$ and
$\delta g_{Af}$ for $f=\nu, l, u,
d$ (assuming universality). Non universal effects (i.e. b quark terms
different from s quark terms)
already appear within SM because of large
$m^2_t$ effects in $Zb \bar b$ couplings\cite{mt2}.
NP can add further non
universal terms which can be described by eq(50). This leads us to
separately discuss the various subsectors.\par

a)\underline{charged leptonic processes} $e^+e^- \to Z \to l^+l^-$\par
 It is convenient to embed the two parameters $\Delta g_{Vl}$
and $\Delta g_{Al}$ into two gauge invariant parameters, namely
$\epsilon^l_1$ and $\bar s^2_l$. They are precisely defined
through\cite{AB}
\bq \epsilon^l_1 = \epsilon_1 = \delta^{s.e.} - 4\Delta g_{Al}  \eq

\bq \bar s^2_l=s^2_1(1+\Delta \bar\kappa'_l) \eq
\bq \Delta \bar\kappa'_l)={1\over 2s^2_1}(\Delta g_{Vl}-(1-4s^2_1)
\Delta g_{Al}) \eq
and can be  experimentally measured through two "good" observables
\bq \Gamma_l = {G_{\mu}M^3_Z\over24\pi\sqrt{2}}[1+\epsilon^l_1]
[1+(1-4 \bar s^2_l)^2]\eq
and
\bq A_l = {2(1-4 \bar s^2_l)\over1+(1-4 \bar s^2_l)^2} \eq
that can be measured through the polarized asymmetry $A_{LR}=A_l$ (or
through the $\tau$ asymmetry) or through the unpolarized
forward-backward asymmetry $A_{FB,l}={3\over4}A^2_l$.

b)\underline{light quark processes} $e^+e^- \to Z \to q \bar q$\par
We assume universality for the first two families, i.e. $q=u$ or $c$
and $q=d$ or $s$. In this case
one obtains 4 parameters (the
generalization to the non universal case with 8 parameters
can be done in a
straightforward manner). The four parameters  $\Delta g_{Vu,d}$
and $\Delta g_{Au,d}$ are now replaced by the gauge invariant
ones\cite{vertex}
\bq   \epsilon^{u,d}_1 = \epsilon_1 + \delta^{(1)}_{u,d}  \eq
\bq   \bar s^2_{u,d}= s^2_1(1+\Delta\bar\kappa'_{u,d})  \eq
with the parameters  $\delta^{(1)}_{u,d}$ and $\delta'_{u,d}$
 describing
the differences with respect to the leptonic case\par
\bq  \delta^{(1)}_{u,d} =4[\delta g_{Al}\pm\delta g_{Au,d}]\eq
\bq  \Delta\bar\kappa'_{u,d}=\Delta\bar\kappa'_l + \delta'_{u,d}\eq
\bq  \delta'_u=-{1\over2s^2}[\delta g_{Vl}-v\delta g_{Al}+
{3\over2}\delta g_{Vu}-({3\over2}-4s^2)\delta g_{Au}]\eq
\bq  \delta'_d=-{1\over2s^2}[\delta g_{Vl}-v\delta g_{Al}-
3\delta g_{Vu}-(3-4s^2)\delta g_{Ad}]\eq
They could in principle be determined by the four observables
 $\Gamma_{u,d}$ or $\Gamma_{c,s}$ and $A_{u,d}$ or $A_{c,s}$.\par
In practice the situation is slightly less simple as one can measure in
an accurate way only $\Gamma_4$ or the combination $D$ given in
eq.(4) and at a weaker level maybe also $\Gamma_c$.\par
 Asymmetry factors $A_q$ are involved in the forward-backward
asymmetries $A_{FB,q}={3\over4}A_lA_q$ but can only be measured
with a sufficient accuracy through polarized $e^{\pm}$ beams
with $A^{pol(q)}_{FB}= {3\over4}A_q$ for q=c and at a weaker accuracy
for q=s.\par
c)\underline{Heavy quark sector}\par
The only process available at Z peak is $e^+e^- \to Z \to b \bar b$.
It contains two additional
parameters \cite{CRV}
that we identify through the departures from universality
with the two first families:
\bq   \delta g_{Vb} = \delta g_{Vd} + \delta g^{Heavy}_{Vb} \eq
\bq   \delta g_{Ab} = \delta g_{Ad} + \delta g^{Heavy}_{Ab} \eq
that can be determined through the two new observables
\bq \Gamma_b = \Gamma_d[1+\delta_{bV}] \eq
\bq A_b = A_d[1+\eta_{b}] \eq
where the coefficients correspond to
\bq  \delta_{bV}=-{4\over 1+v^2_d}[v_d\delta g^{Heavy}_{Vb}
+\delta g^{Heavy}_{Ab}] \eq
\bq  \eta_{b}=-{2(1-v^2_d)\over v_d(1+v^2_d)}[\delta g^{Heavy}_{Vb}
-v_d\delta g^{Heavy}_{Ab}] \eq
with $v_d=1-{4\over3}s^2_1$.
They are measurable through
\bq   R_b = {\Gamma_b\over \Gamma_{had}} \eq
and
\bq   A^{pol(b)}_{FB} = {3\over4}A_b \eq
Note that the parameter $\epsilon_b$ introduced in \cite{AB}
corresponds to a
restricted scheme in which only pure left-handed effects appear
\bq \delta g^{Heavy}_{Vb}=\delta g^{Heavy}_{Ab} ={\epsilon_b\over2} \eq

d)\underline{W mass}\par
The analysis of these precision tests often uses an additional
 observable,
the $W$ mass. This defines one more parameter that is taken as
$\delta\xi$ \cite{BRV} or $\Delta r_{ew}$ or $\epsilon_2$ \cite{AB},
\bq {M^2_W\over c^2M^2_Z} = 1 +\delta\xi \eq
\bq \delta\xi=-{s^2\over c^2-s^2}\Delta r_{ew}  \eq
\bq  \Delta r_{ew} = -{c^2\over s^2}
\epsilon_1+2\epsilon_2+{c^2-s^2\over s^2}\epsilon_3 \eq
One can check that these combinations are vertex correction independent.
At this point
it may be useful for the reader to have a look at Table 3.\par

\begin{center}
\underline{Table 3: Dictionary for universal vacuum polarization
terms
}\par
\vspace{0.2cm}

%
\begin{tabular}{|c|c|c|c|} \hline
\multicolumn{1}{|c|}{Ref.\cite{AB}} &
  \multicolumn{1}{|c|}{Ref.\cite{peskin}} &
   \multicolumn{1}{|c|}{Ref.\cite{old}} &
     \multicolumn{1}{|c|}{Vac. pol.} \\[.1cm] \hline
 $\epsilon_1$ & $\alpha T$ & $\Delta \rho$ & ${A_{33}(0)-A_{11}(0)
\over M^2_W}$
 \\[.1cm] \hline
  $\epsilon_2$ & ${\alpha S\over4s^2}$ & $-c^2\Delta_{3Q}$ &
 ${c\over s}F_{30}(M^2_Z)$
 \\[.1cm] \hline
  $\epsilon_3$ & $-{\alpha U\over4s^2}$ & $\Delta_{1Q}-c^2\Delta_{3Q}$ &
$F_{11}(M^2_W)-F_{33}(M^2_Z)$
 \\[.1cm] \hline
\end{tabular}
\end{center}
\noindent

This table
contains a dictionary for the various notations which had been
introduced in the past for the leading vacuum polarization (universal
or "oblique")
contributions written as\bq \Pi^{ij}(q^2) = A{ij}(0)+q^2F_{ij}(q^2) \eq
Let us also recall the definition
\bq  \Delta \alpha = F_{\gamma\gamma}(0)-F_{\gamma\gamma}(M^2_Z) \eq
and notice one recent notation\cite{Dieter}
\bq  \Delta x=\epsilon_1-\epsilon_2 \ \ \ \Delta y=-\epsilon_2 \ \ \
\epsilon=-\epsilon_3 \eq
We emphazize that the inclusion of
non universal SM or NP terms requires the use of the more general
parametrization defined above with at least 7+1 free parameters.\par

\underline{Brief summary of LEP1 constraints}\par
Within a pure SM analysis, limits on $m_t$ and $M_H$ were obtained from
the sensitivity of the radiative correction
terms to these masses
(essentially the $m^2_t$ and $Log M_h$ dependences). The
sensitivity to $m_t$ is large and has allowed to get a strong constraint

\bq m_t = 177^{+11+18}_{-11-19} GeV \eq
(the first error is experimental, the second one corresponds
to the unknown Higgs mass effect that
is varied between 60 GeV and 1 TeV)
which is in perfect agreement with the observations made at
Fermilab \cite{top}, \cite{mtopD0}
\bq  m_t = 174\pm10^{+13}_{-12} GeV\eq
The bosonic contribution to SM radiative corrections is mainly
concentrated in the parameter $\Delta y$ \cite{Dieter}.
The accuracy is however not sufficient to significantly constrain
 the Higgs mass \cite{Ditt},
although
low values seem to be
favored \cite{LEP1}.\par
In a more general non-standard analysis one can eliminate the
unknown SM
parameters\cite{BRV}
and get constrains on the deviations from the standard $Zf \bar f$
couplings\cite{Zprime}:\par
\noindent
in the leptonic sector (at two standard deviations)\par
\bq \delta g_{Al}\leq 0.002 \ \ \ \ \ \ \delta
g_{Vl}\leq 0.006\eq
\bq  \delta g_{A\nu}=\delta g_{V\nu}\leq 0.005 \eq
in the light quark sector\par
\bq {4\over23}[3 \delta g_{Vu}+9\delta g_{Au}-6\delta g_{Vd}
+4\delta g_{Ve}+23\delta g_{Ae}]\leq 0.008\eq
and in the heavy sector\cite{RV}\par
\bq  \delta_{bV}=0.0414\pm0.0110 \eq
This last result is obtained
for $m_t=175 GeV$, and it constitutes
the first sign for a possible departure from
SM predictions.\par

So \underline{in conclusion}
no NP effect appears in the light fermion sector at an
accuracy reaching a few permille. Some effect may exist in the heavy
quark sector at an accuracy of a few percent.\par
{\bf B)\underline{Indirect constraints on the bosonic sector}}\par
We are now able to discuss the influence of the 11 effective
lagrangians describing NP effects in the bosonic sector.\par

Non blind operators contribute directly (at tree level) to the
$\epsilon_i$ parameters of the light
fermionic sector. Consequently the constraints on the coupling
constants are very strong\cite{DeR}, \cite{Hag}, i.e.
\bq  |\bar f_i{M^2_Z\over\Lambda^2}|
\lsim O(10^{-2}\ \ to\ \ 10^{-3}) \eq
note nevertheless that there is not enough information to disentangle
all possible contributions because of strong correlation effects.\par
Blind operators affect the LEP1 parameters only at 1-loop. The use of
effective lagrangians for loop computations has raised a lot of
technical and physical questions, especially because of the occurence
of strong divergences in some cases\cite{DeR}, \cite{Hag}.
The physical meaning of
these divergent terms is that the chosen effective lagrangian does not
sufficiently specify the NP effects when $q^2$ approaches $\Lambda^2$.
The model has to be completed by additional terms. Restricting to terms
involving only usual particles, the gauge invariance prescription is a
(non unique) way of choosing such additional terms (multi-boson terms,
terms involving Higgs bosons,...). In this case the cancellation of
the violent divergences is provided by diagrams involving the
additional 4-boson couplings and/or the ones involving Higgs bosons.
These features illustrate the model-dependence of these indirect effects.
Another technical point is the fact that the domain of
integration
corresponding to the divergent part may correspond to
a strong coupling regime (and even overpass
the unitarity limit) so that non-perturbative effects should in
principle be taken into account. This weakens the power of the
constraints that has been derived from perturbative analyses.
Nevertheless they give an orientation. In any case, because of the
loop factor ${\alpha \over {4\pi}}$ the constraints are
much weaker than in the case of nonblind operators
\bq  |\bar f_i{M^2_Z\over\Lambda^2}| \lsim O(1\ \ to \ \ 10^{-1})\eq
for example\cite{Hag}\par
\bq  |\lambda_W| \lsim 0.6  \eq
When more than one operator at a time is considered, again because of
strong correlation effects no useful
constraint remains.\par
A very special case has however been noticed \cite{RV}. It has been
shown that the $Zb\bar b$ width allows to get a rather unambiguous
constraint on the $O_{W\Phi}$ operator (with coupling constant $f_W$),
and (owing to specific counting factors and other
numerical factors) a negligible effect of the $O_{B\Phi}$ one (with
coupling constant $f_B$). This arises because
of the
$m^2_t$ enhancement factor which selects
the longitudinal modes of the $W$
couplings inside the loop.

\bq -0.40 \lsim [\bar f_W{M^2_Z\over2\Lambda^2}-
0.04\bar f_B{M^2_Z\over2\Lambda^2}] \lsim -0.15 \eq
The present experimental result which seem to indicate a non zero NP
effect, if it is interpreted as a bosonic effect, would imply a rather
strong departure to SM,
\bq -0.7 \lsim \delta_Z \lsim -0.3 \eq
largely visible in direct tests at LEP2 (that
we shall discuss in the next Sect.5).\par
In addition, if no anomalous effect is  simultaneously observed in the
light fermion sector ($\epsilon_i$ parameters), this means that
 the
$\O_{W\Phi}$ contribution has to cancel against
other contributions to these parameters.
In turn this implies an even richer set of observable effects at LEP2
due to these other sources .\par
On another hand one can compare the order of magnitude of the indirect
constraints obtained in this way with purely
theoretical considerations like the
unitarity relations that we mentioned earlier \cite{GLRUNI}.
It appears that these LEP1 indirect tests can only feel
effects associated to a scale which is weaker than 1 TeV
(after all this not so surprizing for 1-loop effects at Z peak).
However, if the effective scale is only of a few hundreds
of GeV, the validity of the pure perturbative treatment is
questionable, and it is not reasonable to take these numerical
values too
strictly.\par
 The lesson of this discussion
is that only direct tests can give unambiguous
results and this is what we shall discuss in the next Sect.5.\par

\newpage
\section
{Tests at future machines}\par
The future machines that we shall consider are LEP2, LHC and NLC. There
exist also projects for developing LEP1 (Polarized beams, high
luminosities) and extending the Tevatron energy to 4
TeV. With these developments and these new machines one can expect to
improve the indirect tests\cite{str}, \cite{fourf},
but the real progress in the
empirical knowledge of the bosonic sector will only come from a copious
production of boson pairs.\par
Up to now only very mild direct limits have been obtained at
CERN\cite{CERNkappa} and at
Fermilab\cite{FNALkappa}, with anomalous $\kappa$ or $\lambda$ of the
order of the unity.
The first really significant results should come from LEP2 when a few
thousands of $e^+e^-\to W^+W^-$ events will be
observed\cite{BMTlep2}. The standard
reactions $e^+e^- \to ZZ$, $Z\gamma$, $\gamma\gamma$ do not involve
$VW^+W^-$ couplings. Purely neutral non-standard
3-boson couplings $\gamma ZZ$,
$\gamma \gamma Z$, $ZZZ$ may exist but their effects are expected
to be depressed \cite{boudj}.
If by chance
the Higgs boson is light enough to be produced through $e^+e^- \to ZH$
or
$e^+e^- \to \gamma H$, the first meaningful
tests of Higgs couplings could also be
performed\cite{GLRhiggs}.\par
More possibilities will then be offered at LHC\cite{oldLHC}.
Through quark-antiquark
annihilation one can also produce $W^+W^-$, $ZZ$, $Z\gamma$,
$\gamma\gamma$
neutral pairs, but the first new feature is the existence of charged
pair production $W^{\pm}Z$ and $W^{\pm}\gamma$ through $W^{\pm}$
exchange diagram, which will allow to disentangle anomalous $ZW^+W^-$
couplings from $\gamma W^+W^-$ ones. Boson-boson fusion processes will
also take place and give genuine new informations involving 4-boson
couplings and Higgs exchanges\cite{GLRlhc}.\par
At a linear $e^+e^-$ collider (NLC) in the TeV range \cite{physnlc}
the same processes
already studied at LEP2 will be pursued at higher energies and with a
higher luminosity\cite{BMTnlc},\cite{GLRnlc}.
Boson-boson fusion processes will also appear \cite{bb},
\cite{Schrempp}. A
very appealing way to observe this set of processes is through laser
induced \cite{laser} photon-photon collisions
and also photon-electron collisions.
They should be especially interesting for direct Higgs production
($\gamma\gamma \to H$)\cite{GLRhiggs}.\par
In all these direct observations of the bosonic sector,
strategies have to
be developed in order to identify the nature of a possible anomalous
effect or to give sensible observability limits. An observation means a
departure from the SM prediction in a given process. The sensitivity of
any observable to an NP effect generally behaves like
\bq    f({s\over \Lambda^2})^n  \eq
This applies to production rates, ratios of cross sections,
angular distributions, polarization asymmetries,...\par
Particularly
interesting cases are those where the SM contribution is depressed
(for example when it occurs only at 1-loop like in $\gamma\gamma H$ or
in $\gamma\gamma ZZ$) so
that any signal would be a candidate for NP.\par
Let us just mention a few highlights extracted from
the phenomenological
studies that have been
recently made in these processes.\par
In  $e^+e^-\to W^+W^-$, precision tests require the analysis of the
final $W^{\pm}$ polarization. The separation of $W_TW_T$,
$W_LW_L$, $W_TW_L$ production allows to disentangle the 3 types of
C and P conserving
anomalous couplings $\delta$, $\kappa$ and $\lambda$\cite{BMTlep2},
\cite{BMTnlc}, see Fig.7,8 of \cite{BMTlep2}.
The anapole couplings lead to strong
forward-backward asymmetries. The CP violating couplings can be
isolated by doing an analysis of the $W^{\pm}$ spin density matrices
measurable through their decay distributions \cite{CP}.\par
The reaction  $e^+e^-\to \gamma H$ is only observable if it is enhanced
by
anomalous Higgs couplings, for example those
generated by the operators $\O_{UB}$ and $\O_{UW}$\cite{GLRhiggs}.\par
At LHC many processes with different initial states overlap and it will
be difficult to identify the origin of an effect. In a restricted case
with only $\O_W$ and $\O_{UW}$ involved, it has been shown that
ratios of cross sections like ${\sigma(WZ)\over \sigma(ZZ)}$ or
${\sigma(W\gamma)\over \sigma(ZZ)}$ allow a clear disentangling
of $O_W$ and of
$O_{UW}$ effects\cite{GLRlhc}. See Fig.7 of \cite{GLRlhc}.\par
Finally we quote the laser induced
$\gamma\gamma \to H$ process which can give
the highest sensitivity to anomalous Higgs couplings\cite{GLRhiggs}
see Fig.1 of \cite{GLRhiggs}.\par

Details about these preliminary analyses can be found in the
quoted references. Further
more elaborate studies are in progress \cite{network}.

\newpage
\section{Perspectives}\par

The results of the preliminary analyses which have been done along the
lines presented above are summarized in Table 4 and 5.\par

\begin{center}
\underline{Table 4:  Observability limits for $\lambda_W\O_W$ }\par
\vspace{0.2cm}
\begin{tabular}{|c|c|c|c|} \hline
\multicolumn{1}{|c|}{Collider} &
  \multicolumn{1}{|c|}{$|\lambda_W| \lsim$} &
   \multicolumn{1}{|c|}{$\Lambda_{sat} \gsim$} &
     \multicolumn{1}{|c|}{Reference} \\[.1cm] \hline
 LEP2 170GeV & 0.14 & 0.9TeV & \cite{BMTlep2} \\
 LEP2 230GeV & 0.06 & 1.4TeV & \cite{BMTlep2} \\ \hline
 LHC & 0.01 & 3.5TeV & \cite{GR2} \\ \hline
 NLC 0.5TeV & 0.008 & 4TeV  & \cite{BMTnlc} \\
 NLC 1TeV &  0.002 & 8TeV & \cite{BMTnlc} \\ \hline
\end{tabular}
\end{center}
\noindent
\begin{center}
\underline{Table 5:  Observability limits for $d\O_{UW}$ }\par
\vspace{0.2cm}
\begin{tabular}{|c|c|c|c|} \hline
\multicolumn{1}{|c|}{Collider} &
  \multicolumn{1}{|c|}{$|d| \lsim$} &
   \multicolumn{1}{|c|}{$\Lambda_{sat} \gsim$} &
     \multicolumn{1}{|c|}{Reference} \\[.1cm] \hline
 LHC (WW) & 0.1 & 2.5TeV & \cite{GR2} \\  \hline
 NLC (WW) & 0.25-0.02 & 1.5-6TeV  & \cite{GLRnlc} \\
 NLC (HZ) &  0.005 & 11TeV & \cite{GLRnlc} \\  \hline
 laser NLC (H) & 0.001 & 30TeV & \cite{GLRhiggs} \\ \hline
\end{tabular}
\end{center}
\noindent

In these tables we show the sensitivity
to two typical $SU(2)_c$ conserving bosonic couplings,
the anomalous 3-gauge boson
coupling $\lambda_W$ associated to the operator $\O_W$ and the anomalous
Higgs boson coupling $d$ associated to the $\O_{UW}$ operator.\par
These results can be compared to the present indirect LEP1
constraints\cite{Hag}
\bq   |\lambda_W| \lsim 0.6 \ \ \, \ \ \ |d| \lsim1. \ \ \,\ \eq
and to the unitarity bounds for $\Lambda =
1TeV$ \cite{GLRuni}, \cite{GLRUNI}
\bq |\lambda_W| \lsim 0.12 \ \ \, \ \ \ |d| \lsim0.3 \ \ \,\ \eq

\vspace{0.3cm}

\underline{In conclusion}, with these analyses
one observes that step by step the sensitivity will increase when going
from LEP2 to LHC and to NLC, reaching finally the $10^{-3}$ level of
accuracy.  So at the
end the bosonic sector should be tested at the same accuracy as the
fermionic sector is tested at Z peak.
In terms of NP scale as shown in Tables 4,5
this means an order of magnitude of
about 10 TeV. This range of scales is
interesting because it covers a domain in which
several types of theoretical models predict NP effects.\par
It is also exciting to follow the way these progress may arise:\par
---From the high precision direct tests of the fermionic
sector at LEP1, one gets
indirect hints about the gauge boson sector. \par
---The next step starts
at LEP2 with direct tests of the gauge boson sector and some
indirect hints about the Higgs sector.\par
---Finally at LHC, and better at NLC,
direct tests of the Higgs sector should be achieved.\par
\vspace{0.3cm}
Should they give
some indirect hints about a possible underlying
sector at the origin of the mass
generation mechanism?!\par

\end{document}